# Evolutionary Games and Computer Simulations

Bernardo A. Huberman and Natalie S. Glance

Dynamics of Computation Group
Xerox Palo Alto Research Center
Palo Alto, CA 94304

## Abstract

The prisoner's dilemma has long been considered the paradigm for studying the emergence of cooperation among selfish individuals. Because of its importance, it has been studied through computer experiments as well as in the laboratory and by analytical means. However, there are important differences between the way a system composed of many interacting elements is simulated by a digital machine and the manner in which it behaves when studied in real experiments. In some instances, these disparities can be marked enough so as to cast doubt on the implications of cellular automata type simulations for the study of cooperation in social systems. In particular, if such a simulation imposes space-time granularity, then its ability to describe the real world may be compromised. Indeed, we show that the results of digital simulations regarding territoriality and cooperation differ greatly when time is discrete as opposed to continuous.

Over the past decade, the increasing speed and availability of powerful computers has made computer simulations an attractive method to research the behaviours of complex systems. In the physical sciences, simulation techniques have been used to study problems such as critical phenomena, dynamical systems and the large scale structure of the universe. In chemistry and biology, computer simulations provide insight into the mechanics of protein folding and cell metabolism. Similarly, the appearance of behavioral patterns in social settings can be studied using this method, provided that the assumptions of the model capture the underlying interactions.

In a recent paper, Nowak and May (1) presented a set of intriguing results concerning the evolution of cooperation among players placed on a two-dimensional array and confronted with a prisoner's dilemma, which in recent years has become a metaphor for the evolution of cooperation. By running a number of computer simulations, they showed that when players interact with their neighbours through simple deterministic rules and have no memory of past events, the overall evolution produces striking spatial patterns, in which cooperators and defectors both persist indefinitely. Furthermore, for certain parameter values, they observed that regardless of initial conditions, the frequency of cooperators always reaches the same proportion, raising the interesting issue of the existence of a universal constant governing prisoners' dilemma interactions on a lattice. These results were further elaborated on by Sigmund (2), who used them as evidence that territoriality favours cooperation among biological organisms and also suggested that similar results would occur in the case of stochastic transition rules.

While it has been known for some time that cellular automata with deterministic rules can generate pleasing spatio-temporal patterns (3), their usefulness for studying real world systems is not straightforward. One reason is that the granularity imposed by cellular digital machines on both the spatial and temporal domains can generate behaviors that may not have counterparts in the continuum limit.

In fact, there are important differences between the way a system composed of many interacting elements is simulated by a digital machine and the manner in which it behaves when studied in real experiments. In some instances, these disparities can be marked enough so as to cast doubt on the implications of cellular automata type simulations for the study of cooperation in social systems. These differences, which have been analyzed in detail for simulations of Ising-like magnets and other condensed matter systems (4,5,6), also have important implications for computer studies of evolutionary games. This issue must be addressed if computer simulations are to provide insight into social and biological dynamics.

We begin by analyzing the discrepancies between some digital simulations and real systems and afterwards compare the results of running the same computations presented by Nowak and May in both synchronous and asynchronous fashion. Our results for asynchronous updating demonstrate that defection is the dominant evolutionary outcome if at least one defector is present in the initial configuration. The matrix converges rapidly to steady-state and does not display the widely changing spatial patterns observed in the synchronous case. These results, which do not correspond at all to the behavior found for synchronous updating, cast into doubt the conclusions recently obtained concerning territoriality and the universality of long-term averages of cooperation.

Before considering the particular computer experiments studied by Nowak and May, it is useful to clarify the differences between a cellular automata simulation of a natural process and the dynamics of the same system as found in nature. In a simulation of the type presented by the authors of Ref. 1, the general computation goes through a series of discrete states which are updated by the program at integral values of unit time, according to a set of given instructions. Nothing happens for times shorter than this unit time. The simulations presented in Ref. 1 are *synchronous*, meaning that all the players are updated in unison at every time step. The resulting global dynamics is mathematically described by a finite difference equation, which if sufficiently nonlinear can generate the complicated patterns and chaotic outcomes that are displayed in their paper.

In natural social systems, however, a global clock that causes all the elements of the system update their state at the same time seldom exists. While clocks and seasonal effects can synchronize metabolic and reproductive processes in biological organisms, in most social settings, players, agents, or organisms act at different and uncorrelated times on the basis of information that may be imperfect and delayed. At the same time that one player initiates an interaction with its neighbor, another pair may be wrapping up an ongoing game. In the physical realm, the interactions among the individual atoms or molecules making up a magnet or a crystal determines the behaviour of the material without there being a global clock which synchronizes their actions. In this *asynchronous* world, the dynamics is usually expressed in the form of differential equations, whose solutions are not always the same as those of their finite difference counterparts

This analysis implies that if a computer simulation is to mimic a real world system with no global clock, it should contain procedures that ensure that the updating of the interacting entities is continuous and asynchronous. This entails choosing an interval of time small enough so that at each step at most one individual entity is chosen at random to interact with its neighbours. During this update, the state of the rest of the system is held constant. This procedure is then repeated throughout the array for one player at a time, in contrast to a synchronous simulation in which all the entities are updated at once.

In order to show the striking differences between synchronous and asynchronous simulations of cooperative games, we conducted a number of computer experiments using the same scenario presented by Nowak and May. Fig. 1(*a*) shows the results of a computer experiment with synchronous updating of a system of two dimensional players on a $99 \times 99$ square-lattice world with fixed boundary conditions after 217 generations. The initial condition consists of a single defector at the center surrounded by a world of cooperators. As can be seen, the resulting symmetrical pattern is the same as that shown by Nowak and May in Fig. 3(*c*) of their paper.

The evolution of the same system is very different when the players' strategies are updated asynchronously. In our experiments, we choose the time interval between updates to be small enough so that at most one player updates within that time interval. As a consequence of this asynchronous updating, each player awakens to see a slightly different world than the players acting before or afterwards. The updating is performed as follows. Each player receives a score per unit time that is updated continuously as the matrix of players varies over time. In analogy to the discrete scenario considered in the synchronous case, each player's score per unit time is the sum of its payoffs per unit time. These payoffs are received from interactions with each of its neighbors and itself in a prisoner's dilemma confrontation.

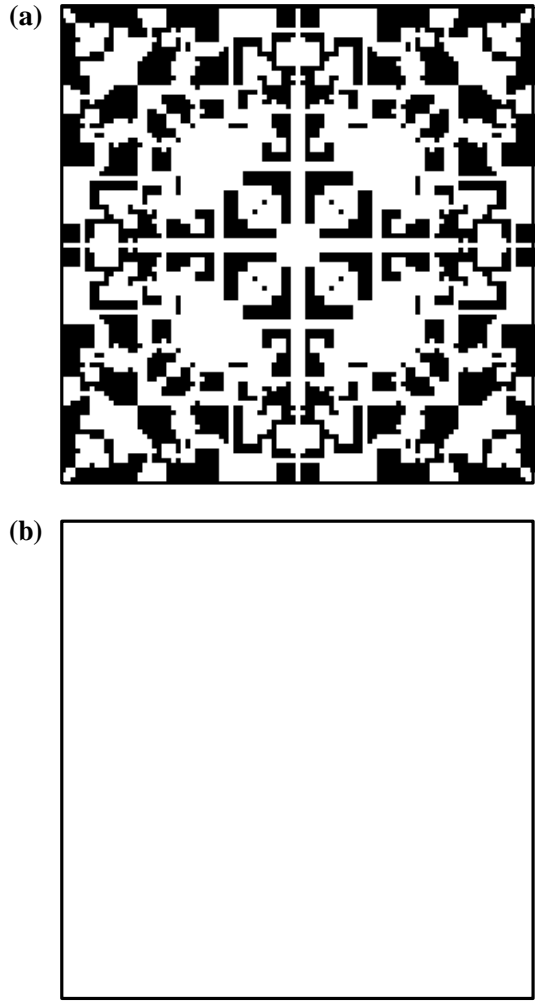

**Fig. 1.** *Synchronous versus asynchronous updating* of player actions results in differing evolutionary pathways. This simulation starts with a single defector at the center of a 99 x 99 square lattice world of cooperators with fixed boundary conditions. The coding is as follows: black represents a C site and white represents a D site. (a) Player actions are updated synchronously. This snapshot was taken at generation $t = 217$ and corresponds to Fig. 3(b) of Nowak and May's paper. (b) Player actions are updated asynchronously. Within a hundred generations or so, the matrix evolves into a fixed state in which all of the players are defecting.

Next, within a microstep, at most one player is replaced by the highest scoring player within its neighborhood. The size of the microstep is chosen so that the average updating time (equivalent to one generation) for the whole array is the same for both the synchronous and asynchronous cases.

Fig. 1(*b*) shows the asynchronously updated system at the same point in time as that of Fig. 1(*a*), starting from identical initial conditions. Within a hundred generations or so, the array evolves into a fixed state in which all of the players are defecting. In fact, as long as there is at least one defector in the initial state, the matrix always evolves rapidly into a state of overall defection.

Alternatively, one might hypothesize that synchronous simulations of evolutionary games are relevant to real world systems in which there are delays in the transmission of information. Delays would then cause player states to be updated on the basis of neighbourhood configurations that correspond to earlier times. Indeed, if the player's score were intended to reflect its fitness function (*i.e.*, its ability to reproduce), one would expect a player's present score (or number of offspring) to depend on past interactions, as well as present ones.

However, in cases where player interactions depend upon delayed information, a simulation may still possess many complicated dynamical features that will not reduce to the case of synchronous updating. In our experiments, we found that when the players are updated asynchronously based on delayed scores, the evolution of cooperation depends strongly on the size of the delay: the greater the delay the higher the asymptotic level of cooperation. In addition, for very large delays (corresponding to several generations), we observed initial transients that lengthened as delays increased.

These results show that while computer experiments provide a versatile approach for studying complex systems, an understanding of their subtle characteristics is required in order to reach valid conclusions about real world systems. The instance provided by Nowak and May is only one of many cases where the outcomes of synchronous evolutionary games on computers have been invoked to provide insights into the workings of living systems. Other examples are provided by computer experiments used to validate theories about the emergence of order in evolution (7), phenotypic novelties through progressive genetic change (8), the molecular origin of life (9) and studies of artificial life forms through computer simulations (10). Unless it can be demonstrated that global clocks synchronize mutations and chemical reactions among distal elements of biological structures, the patterns and regularities observed in nature will require continuous descriptions (11) and asynchronous simulations.